\documentclass[aps,10pt,prb,twocolumn,showpacs,amsmath,amssymb]{revtex4-1}

\usepackage{graphicx}
\usepackage{ifthen}
\usepackage{color}
\usepackage{bm} 
\usepackage{multirow}
\usepackage{dcolumn}
\usepackage{bbold} 
\usepackage{extarrows}
\usepackage{hyperref}
\hypersetup{
 pdfnewwindow=true, colorlinks=true,
 linkcolor=blue, anchorcolor=blue,
 citecolor=blue, filecolor=blue,
 menucolor=blue, urlcolor=blue}
\usepackage{breakurl}

\newcolumntype{.}{D{.}{.}{-1}}

\begin{document}

\title{Theoretical study of solid iron nanocrystal movement inside a
  carbon nanotube}

\author{Sinisa Coh}
\email{sinisa@civet.berkeley.edu} 
\author{Steven G. Louie} 
\author{Marvin L. Cohen} 
\affiliation{Department of Physics, University of California at Berkeley,
  Berkeley, CA 94720, USA}
\affiliation{Materials Sciences Division, Lawrence Berkeley National
  Laboratory, Berkeley, CA 94720, USA}

\date{\today}


\begin{abstract} 
  We use a first-principles based kinetic Monte Carlo simulation to
  study the movement of a solid iron nanocrystal inside a carbon
  nanotube driven by the electrical current. The origin of the iron
  nanocrystal movement is the electromigration force. Even though the
  iron nanocrystal appears to be moving as a whole, we find that the
  core atoms of the nanocrystal is completely stationary, and only the
  surface atoms are moving. Movement in the contact region with the
  carbon nanotube is driven by electromigration forces, and the
  movement on the remaining surfaces is driven by diffusion. Results
  of our calculations also provide a simple model which can predict
  the center of mass speed of the iron nanocrystal over a wide range
  of parameters.  We find both qualitative and quantitative agreement
  of the iron nanocrystal center of mass speed with experimental data.
\end{abstract}

\maketitle

\section{Introduction and motivation}

The interior of multiwall carbon nanotubes can be filled with various
metallic nanocrystals. Additionally a metallic nanocrystal will start
to move inside a carbon nanotube if an electrical current is applied
axially to the carbon nanotube.  The speed of the nanocrystal can be
tuned over many orders of magnitude, since the speed of the
nanocrystal depends exponentially on the applied electrical
current\cite{Begtrup2009}. The motion of the metallic nanocrystal on
the interior or exterior of the carbon nanotube has been observed
previously for
iron\cite{Svensson2004,Begtrup2009,Begtrup2009PRB,Loffler2011},
copper\cite{Golberg2007}, tungsten\cite{Jin2007},
indium\cite{Regan2004}, and gallium\cite{Min2012}. The movement of
nanocrystals inside carbon nanotubes is interesting from the
perspective of memory applications\cite{Begtrup2009}, as a constituted
element of nanomachines, or for tunable synthesis of metal
nanocrystals\cite{CohPRL2013}.

The direction of the nanocrystal movement depends on the polarity of
the applied electrical current. Therefore, the movement of a
nanocrystal most likely originates from electromigration forces
acting on the metallic atoms such as the electron wind force. However, the
precise mechanism of nanocrystal movement is not well understood.
Additionally, recently it was found experimentally\cite{CohPRL2013}
that an iron nanocrystal of a given diameter can move through a
constriction inside a carbon nanotube of a smaller diameter while
remaining an ordered solid.

We performed a series of theoretical calculations to try to understand
the nature of the movement of metallic nanocrystals inside carbon
nanotubes in more detail. We model a nanocrystal of iron,
since this is a commonly studied nanocrystal. However we expect
that the qualitative nature of movement of other metal nanocrystals
will be similar to that of iron. We find that even though it appears
that the iron nanocrystal is moving as a whole through the nanotube,
in fact, the individual iron atoms are only moving on the surfaces of
the nanocrystal. The bulk iron atoms remain stationary as long as they are
in the bulk. Once the atoms that were in the bulk are exposed to the
end surface, they move along the interface with the carbon nanotube
towards the front surface. A somewhat related mechanism, but involving
heating of iron nanocrystals and its chemical reaction with the carbon
nanotube, was proposed in Ref.~\onlinecite{Loffler2011}.

\section{Methods}

Here we describe our theoretical modeling of the movement of an iron
nanocrystal inside a carbon nanotube.

Although density functional theory (DFT) is a powerful technique for
first-principles study of material properties, it is most commonly used
to study systems with stationary positions of atoms. With the help
of a molecular dynamics method\cite{Car1985}, one can study dynamical
properties from first-principles. However, in practice one can use
first-principles molecular dynamics method only on time scales
comparable to period of atomic vibrations $\sim 10^{-12}$~seconds.

In order to study movement of an iron nanocrystal inside a carbon
nanotube, we need to analyze the processes on time scale close to
$\sim 10^{-3}$~seconds since typical energy barriers for iron atom
movement are close to 0.6~eV, while the relevant temperature is about
twenty times smaller, 0.03~eV. Therefore typically an iron atom will
jump across the barrier once in $\sim e^{0.6/0.03} \sim 10^9$ atomic
vibrations, or equivalently, once every $\sim 10^{-3}$ seconds.  In
order to deal with these rare events, we approximate the time dynamics
of this system using the kinetic Monte Carlo (kMC) method.  The
kinetic Monte Carlo method ignores the details of time dynamics,
instead it deals only with fixed crystal sites which do or do not
contain an atom at a given time. For each time step one moves an atom
from one site to the next according to a rate given by energy barrier
height of the move in question. Therefore, kMC simulations require
only knowledge of energy barriers for iron diffusion processes.

In order to obtain reliable energy barriers for iron diffusion, we
perform first principles DFT calculations of selected subsets of
relevant diffusion processes in iron. We find that these barriers
depend strongly on the environment of the iron atom (for example, bulk
diffusion has a larger barrier than surface diffusion). Since it becomes
combinatorially expensive to compute all possible iron diffusion
barriers, we constructed a simple model for an estimation of any given
diffusion barriers which we parametrize using our DFT calculations. We
also incorporated in this model the interaction of iron atoms with the
carbon nanotube. Later we will show that the qualitative nature of our results is
robust under changes of parameters of this simple model.

In our calculations we do not discuss the microscopic origin of
the electromigration forces on iron atoms from the current flowing through the
carbon nanotube. We simply consider the electromigration force per iron atom as
a parameter. Nevertheless, based on our results and experimental data
from Ref.~\onlinecite{Begtrup2009} and \onlinecite{CohPRL2013} we
speculate that the origin of electromigration forces on iron is the electron
wind force and not a direct force, as is the case for most
metals\cite{Sorbello1997}.

\subsection{Density functional theory calculation of iron
  diffusion}

We perform a density functional theory calculation for a body-centered
cubic iron 
diffusion in the following geometries: bulk iron diffusion and
(001) and (011) surface diffusion. We consider both diffusion of
surface vacancies and surface iron adatoms, and we consider the influence
of a carbon overlayer on iron surface diffusion, and include both first
and second neighbor hopping.  Furthermore, we neglect exchange
diffusion processes and only consider diffusion processes in which
a single atom is displaced between initial and final configuration.

As a first step in the computation of diffusion barriers we perform full
structural relaxation of initial and final configurations of the
diffusion process at hand. The only exception is the relaxation of the
carbon layer, since that would introduce additional numerical noise due
to imperfect lattice matching between iron and carbon
lattices. Therefore we only allowed rigid shifts of entire carbon
layers in the direction perpendicular to the iron surface.

In a second step, we perform a nudged elastic band
calculation\cite{henkelman:9901} with three configurations between
initial and final configuration. For the middle of the three
configurations, we use the climbing image method\cite{henkelman:9901}
to obtain a more precise value of energy barrier.

We performed density functional theory calculations using the
SIESTA\cite{siesta} computer package with a vdW-DF2 density
functional\cite{PhysRevB.82.081101}.  This functional includes
non-local van der Waals interaction, which are important for our
calculation since ordinary GGA functionals show almost no binding
between metal surfaces and a carbon layer\cite{Hamada2010}.  For both
iron and carbon, we use norm-conserving pseudopotentials and
a double-zeta polarized basis set. We use a grid cutoff energy of 440~Ry
and an effective $10 \times 10 \times 10$ k-point grid for conventional
2-atom body-centered cubic unit cell of iron. The nudged elastic band part
of the calculation was performed using an ASE\cite{ase} simulation
environment.

\begin{table}
  \caption{\label{tab:barriers} Density functional theory computed
    diffusion barriers in eV for various geometries either with or
    without an additional carbon layer on top of the iron surface. Diffusion
    pathways are always considered between sites closest to the
    surfaces and between closest first (or second) neighbor bcc
    sites. For each process we also show for the diffusing atom the number of
    first neighbor iron atoms in initial ($Z^i_{\rm Fe}$) and final
    ($Z^j_{\rm Fe}$) configuration.}
\begin{ruledtabular}
  \begin{tabular}{l.cc}
\multicolumn{1}{c}{Type of process} 
& 
\multicolumn{1}{c}{Barrier (eV)} 
&
$Z^i_{\rm Fe}$ & $Z^j_{\rm Fe}$
\\
\hline
\multicolumn{2}{l}{Bulk diffusion} 
\\
\quad \quad to first neigh.                   &  0.72  & 7 & 7 \\
\vspace{0.1cm}
\quad \quad to second neigh.                  &  2.72  & 8 & 8 \\
\multicolumn{2}{l}{Adatom diffusion} \\
\quad \quad on (001) surface to second neigh. &  1.32  & 4 & 4 \\ 
\vspace{0.1cm}
\quad \quad on (011) surface to first neigh.  &  0.36  & 2 & 2 \\ 
\multicolumn{2}{l}{Vacancy diffusion} \\
\quad \quad on (001) surface to first neigh.  &  1.23 \footnotemark[1] & 7 & 3 \\
\quad \quad on (001) surface to second neigh. &  1.17 & 4 & 4 \\
\quad \quad on (011) surface to first neigh.  &  0.55 & 5 & 5 \\ 
\vspace{0.1cm}
\quad \quad on (011) surface to second neigh. &  1.74  & 6 & 6 \\
\multicolumn{2}{l}{Vacancy diffusion with carbon layer} \\
\quad \quad on (001) surface to first neigh.  &  1.27 \footnotemark[2] & 7 & 3 \\ 
\quad \quad on (001) surface to second neigh. &  1.15 & 4 & 4 \\ 
\quad \quad on (011) surface to first neigh.  &  0.54 & 5 & 5 \\ 
\quad \quad on (011) surface to second neigh. &  1.64 & 6 & 6  
\end{tabular}
\end{ruledtabular}
\footnotetext[1]{Without saddle point.}
\footnotetext[2]{Asymmetric diffusion path, barrier for this process
  in the opposite direction, $j \rightarrow i$ is 0.40~eV.}
\end{table}

We list results for the DFT diffusion energy barrier calculations in
Table~\ref{tab:barriers}. We find that processes with lowest energy
barriers are adatom diffusion on the (011) surface (0.36~eV), vacancy
diffusion on the (011) surface (0.55~eV), and first neighbor bulk
diffusion (0.72~eV). All three of these processes involve diffusion
between first neighbor sites. Furthermore, in all of these cases
the initial and final atom configurations along the diffusion pathway are
equivalent (symmetric diffusion).

Second neighbor symmetric diffusion both for bulk diffusion and (011)
surface diffusion is about three to four times larger than symmetric
first neighbor diffusion (2.72~eV versus 0.72~eV and 1.74~eV versus
0.55~eV). For this reason, we neglect second neighbor
diffusion and focus only on first neighbor diffusion.

Furthermore, we find that asymmetric processes have different barriers
than symmetric processes. For example, the energy barrier for first
neighbor diffusion in bulk is 0.72~eV, while first neighbor diffusion
on a (001) surface is almost twice as large, 1.23~eV. In a body-centered
cubic crystal, the first neighbor diffusion on the (001) surface must occur
between the top-most surface layer and the next-to-top-most surface layer.
Therefore the difference in binding energy for an iron atom in these two
layers explains the observed increase in first neighbor diffusion on
the (001) surface.

Furthermore, we find that having a single carbon layer (graphene) next
to the iron surface has negligible influence on surface diffusion of
iron atoms. We computed the distance between the iron surface and the carbon layer
to be 2.45~\AA\ for a (001) surface and 3.18~\AA\ for a (011) surface.

\subsection{Model of iron diffusion}
\label{sec:model}

Based on theoretical calculations of iron diffusion barriers, we next
describe a model which will assign an energy barrier to any diffusion
process in iron.

For the diffusion of an iron atom from site $i$ to site $j$ having the
same number of iron and carbon first neighbors at $i$ and $j$ site
(symmetric diffusion) we assign a diffusion energy barrier $E_{i
  \rightarrow j}^{\rm sym}$ as,
\begin{align}
  E_{i \rightarrow j}^{\rm sym} & = a + b Z_{\rm Fe}^i.
  \label{eq:sym_barrier}
\end{align}
Here $Z_{\rm Fe}^i=Z_{\rm Fe}^j$ is the number of first neighbor iron
atoms at either site $i$ (counted before an atom moves from $i$ to $j$) or
site $j$ (counted after an atom moves from $i$ to $j$). We obtain values
of parameters $a$ (0.21~eV) and $b$ (0.071~eV) from a fit to all
first-neighbor symmetric diffusion barriers from
Table~\ref{tab:barriers}. These three processes also have the lowest
diffusion barriers and they include: adatom diffusion on the (011) surface
(0.36~eV), vacancy diffusion on the (011) surface (0.55~eV), and first
neighbor bulk diffusion (0.72~eV). Fitted values for these processes
given by Eq.~\ref{eq:sym_barrier} are reproduced within 0.02~eV (they
are respectively, 0.35~eV, 0.57~eV, and 0.71~eV).

Since we found that the presence of the carbon layer has almost no influence
on symmetric diffusion processes, the energy barrier $E_{i \rightarrow j}$
in Eq.~\ref{eq:sym_barrier} does not depend on the number of carbon
neighbors.

For asymmetric diffusion of an iron atom from site $i$ to site $j$, where
the number of iron neighbors is different at $i$ and $j$, we assign a diffusion
barrier with an additional penalty term accounting for the change in number
of first neighbor atoms as,
\begin{align}
  E_{i \rightarrow j} & = a + b Z_{\rm Fe}^i + \max 
    \left( 0, c \Delta Z_{\rm Fe} + d \Delta Z_{\rm C}  \right).
  \label{eq:asym_barrier}
\end{align}
$\Delta Z_{\rm Fe}$ is the difference in number of first neighbor iron
atoms between sites $i$ and $j$, while $\Delta Z_{\rm C}$ is
difference in the effective number of first neighbor carbon atoms. In
section~\ref{sec:kmc} we describe how we assign the effective number
of first neighbor carbon atoms.

The parameter $c$ quantifies the strength of the interaction between
neighboring iron atoms, and it is formally similar to the exchange $J$
parameter in the Ising model. We obtained the value of parameter $c$
(0.31~eV) by comparing a DFT computed total energy for a structure
with an iron vacancy in the first layer on a (001) surface to
structure with iron vacancy in second layer on (001) surface. When the
iron vacancy is in the first layer, the total ground state energy is
lower by 1.23~eV.  Since in this process exactly four iron-iron first
neighbor pairs get broken ($Z^i_{\rm Fe} - Z^j_{\rm Fe} = 7 - 3= 4$),
we obtain $c=1.23/4=0.31$~eV. We decided not to fit first-neighbor
diffusion process on the (001) surface directly to
Eq.~\ref{eq:asym_barrier}, since in our DFT calculations we find that
this process does not have an energy saddle point, instead the energy
is monotonically increasing while going from the initial to the final
configuration. Instead, we find it more important to obtain a more
reliable value of the iron-iron interaction strength. We obtain a
similar value of parameter $c$ by considering the displacement of the
iron vacancy between first and second layers of the (011) surface
(0.33~eV) where only two iron-iron first neighbor pairs get broken.
Somewhat larger values of parameter $c$ are obtained from surface
formation energy of the (001) surface (0.37~eV) and the (011) surface
(0.50~eV). In section~\ref{sec:robustness} we show the robustness of
our results to changes to value of this and other model parameters.

Parameter $d$ quantifies the strength of the iron-carbon interaction.
We obtained a value for parameter $d$ (0.14~eV) by comparing the DFT
computed energy of an iron surface terminated with a carbon layer
(graphene) to a DFT energy of clean iron surface and a carbon layer in
vacuum. The value for the parameter $d$ for a (001) iron surface
(0.13~eV) is somewhat smaller than on a (011) iron surface (0.15~eV),
which is why we use their arithmetic average in the calculation.

In our model, we neglect diffusion of iron atoms to second nearest
neighbor since those processes have higher diffusion barriers (see
Table~\ref{tab:barriers}). Furthermore, our DFT calculations show that
the energy required to remove single iron atom from bulk or surface to
the vacuum is much underestimated by the penalty term $c \Delta Z_{\rm
  Fe}$ from Eq.~\ref{eq:asym_barrier}.  For such process one would
need to use an effective value of $c$ to be 0.90, 1.05, or 1.46~eV
(for an atom removal from bulk, (011), and (011) surfaces
respectively) which is three to five times larger than the value of
$c$ we obtained earlier. For this reason, we have decided to simply
neglect processes in which an atom moves to site $j$ which does not
have any iron atoms in first neighbor sites ($Z^j_{\rm Fe}=0$).

\subsection{Kinetic Monte Carlo simulation}
\label{sec:kmc}

Given the model from Sec.~\ref{sec:model} to describe the diffusion
process in iron, we can proceed to do a kinetic Monte Carlo simulation
of an iron nanocrystal movement inside a carbon nanotube.

We first define a fixed set of atomic sites along which iron atoms can
move. We start with an infinite arrangement of body-centered cubic
sites with an iron lattice constant $a=0.29$~nm. Next, we construct a
cylinder with radius $r_{\textrm{cyl}}$ about an order of magnitude
larger than $a$ immersed in this infinite arrangement of body-centered
cubic lattice sites.  We ignore all iron sites outside of this
cylinder and consider only sites inside the cylinder, to simulate an
iron nanocrystal contained inside carbon nanotube.  We take the
orientation of the cylinder axis to point along the crystal [100]
direction as found experimentally\cite{Begtrup2009PRB}. At the
beginning of the simulation (time $t=0$), we occupy certain number of
such sites within the tube, while all other sites inside the cylinder
(carbon nanotube) are initially empty. For simplicity we always start
from a configuration in which the occupied sites are taken in a
certain range of heights $[z_{\rm min},z_{\rm max}]$ along the
cylinder axis.

Starting from an initial arrangement of iron atoms we compile a list
of all possible moves that iron atoms can make. Our DFT calculation
has shown that it is sufficient to consider only moves of iron atoms
to empty first nearest neighbor sites. To each such move from the list
(between sites $i$ and $j$) we assign a rate $\rho_{i \rightarrow j}$,
\begin{align}
  \rho_{i \rightarrow j} = \rho_0 \exp \left[ - \frac{E_{i \rightarrow
        j}}{kT} + \frac{\frac{1}{2}({\bm r}_j - {\bm r}_i)\cdot {\bm
        F}_{\rm em}}{kT} \right].
   \label{eq:rate}
\end{align}
Here $\rho_0$ is constant commonly used in kMC modeling
($\rho_0=10^{12}$~s$^{-1}$) corresponding to the inverse of a typical
phonon frequency, $k$ is Boltzmann constant, $T$ is the simulation
temperature, and $E_{i \rightarrow j}$ is the energy barrier height as
computed from a first-principles based model given in
Eq.~\ref{eq:asym_barrier}.

Since an iron nanocrystal and a carbon layer have incommensurate
lattices, the assignment of the number of carbon neighbors to a given
iron site becomes difficult. Therefore we employ the following simple
scheme to assign the number of first neighbor carbon atoms. For each
iron site on the boundary of the nanocrystal, we simply count the
number of first neighbor iron-iron pairs broken by the cylindrical cut
and we assign that number to be the effective number of carbon bonds.

Finally, ${\bm r}_i$ and ${\bm r}_j$ in Eq.~\ref{eq:rate} are
Cartesian coordinate vectors for atomic sites $i$ and $j$, while ${\bm
  F}_{\rm em}$ is the electromigration force acting on an iron atom,
originating from the current in the carbon nanotube. Assuming that the
energy barrier maximum between sites $i$ and $j$ occurs halfway in
between, the factor $\frac{1}{2}({\bm r}_j - {\bm r}_i)\cdot {\bm
  F}_{\rm em}$ appearing in Eq.~\ref{eq:rate} accounts for increase in
diffusion rate $\rho_{i \rightarrow j}$ along the direction of the
electromigration force.  Similarly, this factor reduces diffusion rate
$\rho_{i \rightarrow j}$ in the direction opposite to the
electromigration force.

We assume that the force ${\bm F}_{\rm em}$ is non-zero only if either
site $i$ or $j$ is adjacent to the nanotube (i.e. either site $i$ or
site $j$ has non-zero number of effective carbon neighbors), and we
test robustness of this assumption in section~\ref{sec:robustness}.
Furthermore, we take the vector ${\bm F}_{\rm em}$ to point in the
direction of the cylinder (nanotube) axis, along the direction of the
current. The magnitude of the vector ${\bm F}_{\rm em}$ is then taken
as a parameter of the simulation. In Sec.~\ref{sec:exp_constant} we
relate force ${\bm F}_{\rm em}$ to the electrical current density $j$.

Now that we have assigned the rate $r_{i \rightarrow j}$ to each
possible atomic move $i \rightarrow j$ in the initial configuration,
we proceed by performing atomic steps. We choose which step to perform
based on a standard kMC probabilistic model\cite{Bortz1975,Voter2007}
which chooses at random one of the steps according to the rate given
by Eq.~\ref{eq:rate}. Once the move has been performed the simulation
time is updated from $t=0$ to $t=\Delta$ according to the rates given
by Eq.~\ref{eq:rate}. Since performing this atomic step has altered
the atom configuration, we need to update the list of all possible
moves and update the rates assigned to the moves. With an updated list
of moves and their rates, we repeat this process until we reach the
desired number of simulation steps, or the desired simulation time
$t$. In order to speed up the kMC simulation, we also employ the
binary tree algorithm\cite{Blue1995}.

\section{Results and discussion}

In this section we will present results of our kinetic Monte Carlo
simulation.

\subsection{Character of movement in constant cross-section area
  carbon nanotube}
\label{sec:character}

\begin{figure}
\centering\includegraphics{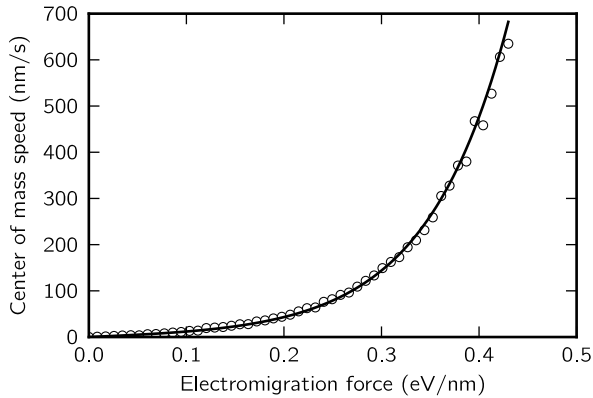}
\caption{Computed iron nanocrystal center of mass speed as a function
  of electromigration force $F_{\rm em}$. Simulation temperature is
  700~K, somewhat larger than in the experiment\cite{Begtrup2009},
  iron nanocrystal radius is $r_{\textrm{cyl}}$=1.05~nm, and the length is
  $l=4.31$~nm. Line is a fit to functional form as in
  Eq.~\ref{eq:fit}.}
\label{fig:exp_regime}
\end{figure}

We find that for a wide range of parameters, the iron nanocrystal can
move through a carbon nanotube under the application of an external
electromigration force. Figure~\ref{fig:exp_regime} shows the
dependence of the nanocrystal center of mass speed on the magnitude of
the electromigration force per atom $F_{\rm em}=|{\bm F}_{\rm em}|$
for fixed simulation temperature, nanocrystal area, and length. It is
clear that the speed depends non-linearly on the external force just
as in the experiments\cite{Begtrup2009,Loffler2011}. We postpone the
analysis of the center of mass speed dependence on force, temperature,
area, and length to sections \ref{sec:length},\ref{sec:tempForce}, and
\ref{sec:area}. Here we first focus on the nature of iron nanocrystal
movement in the nanotube.

It is easier for demonstration purposes to describe the nanocrystal
motion for temperatures somewhat higher than those found in
experiment\cite{Begtrup2009}. Therefore we defer analysis of our model
calculation in experimental range of temperatures to
Sec.~\ref{sec:exp_constant}.

\begin{figure}
\centering
\includegraphics{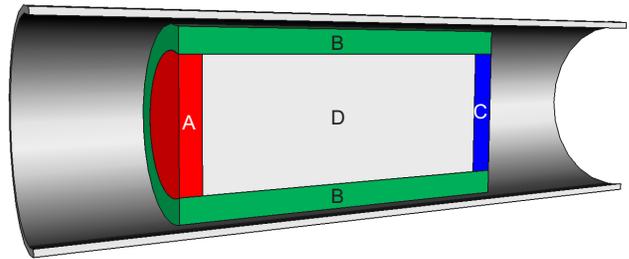}
\caption{Cross-section of iron nanocrystal inside carbon nanotube.
  Four regions of the nanocrystal are indicated (A, B, C, and D) see
  main text for details. For definiteness, the direction of
  the electromigration force is assumed to be to the right in this
  figure.}
\label{fig:3d}
\end{figure}

Figure~\ref{fig:3d} shows the cross-section along cylinder axis of the
iron nanocrystal and carbon nanotube. Four regions of the iron
nanocrystal are indicated.  Regions A, B, and C consist of atoms on
the boundary of the nanocrystal, while atoms in region D are in the
bulk (core) of the nanocrystal. Furthermore, the circular regions A
and C are on opposite sides (caps) of the nanocrystal, while the
cylindrical shell B is in contact with the carbon nanotube. In the
following discussion the assignment of regions A, B, C, and D is
assumed to be stationary, i.e.  atoms can move from one region to
another, but the region assignment relative to nanocrystal remains the
same. For definiteness we assume that the electromigration force is
pointing to the right in Fig.~\ref{fig:3d}.

Our kMC simulation shows that atoms in region D are stationary, as
long as they remain in region D. Atoms in region B under the influence
of the electromigration force get pushed towards region C, where they
diffuse evenly along the cylinder cap. Vacancies created in region B
create a concentration gradient which by diffusion attracts atoms from
region A to region B.

We now focus on the movement of a single atom that starts out in
region A. Under the influence of the diffusion force created by
vacancies in region B, this atom will eventually reach region B. Once
in region B under the influence of the electromigration force it will
move toward region C. Once it reaches region C, it will distribute
there with near uniform probability, again due to diffusion forces.
After more and more atoms get extracted from region A into region C,
this particular atom will eventually get covered by enough layers of
new atoms so that it will effectively become part of region D. Once in
region D, this atom remains stationary! Once all remaining atoms are
removed from region A and added to region C, this atom will become
part of region A and the entire process repeats. Therefore,
schematically the pattern of movement of individual iron atom can be
described as,
\begin{align}
A 
\xlongrightarrow[\textrm{diffusion}]{} 
B
\xlongrightarrow[F_{\rm em}]{}
B
\xlongrightarrow[\textrm{diffusion}]{} 
C
\xlongrightarrow[\textrm{wait}]{} 
D
\xlongrightarrow[\textrm{wait}]{} 
A.
\notag
\end{align}

\begin{figure}
\centering
\includegraphics{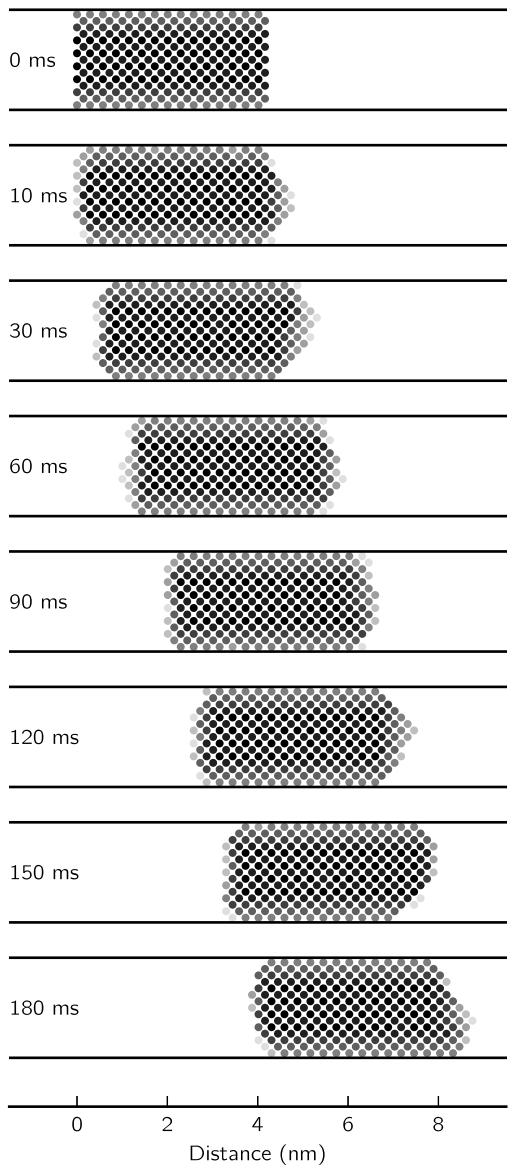}
\caption{Two dimensional projection of three dimensional iron atom
  positions at eight different snapshots in the kinetic Monte Carlo
  simulation.  Various intensities of greyness correspond to rows
  containing more (darker gray) or less (brighter gray) iron atoms.
  The simulation temperature is 700~K, somewhat larger than in the
  experiment\cite{Begtrup2009}, iron nanocrystal radius is
  $r_{\textrm{cyl}}$=1.11~nm, and the length is $l=4.31$~nm.  The
  electromigration force magnitude is $F_{\rm em}=0.28$~eV/nm.}
\label{fig:straight_occ}
\end{figure}

\begin{figure}
\centering
\includegraphics{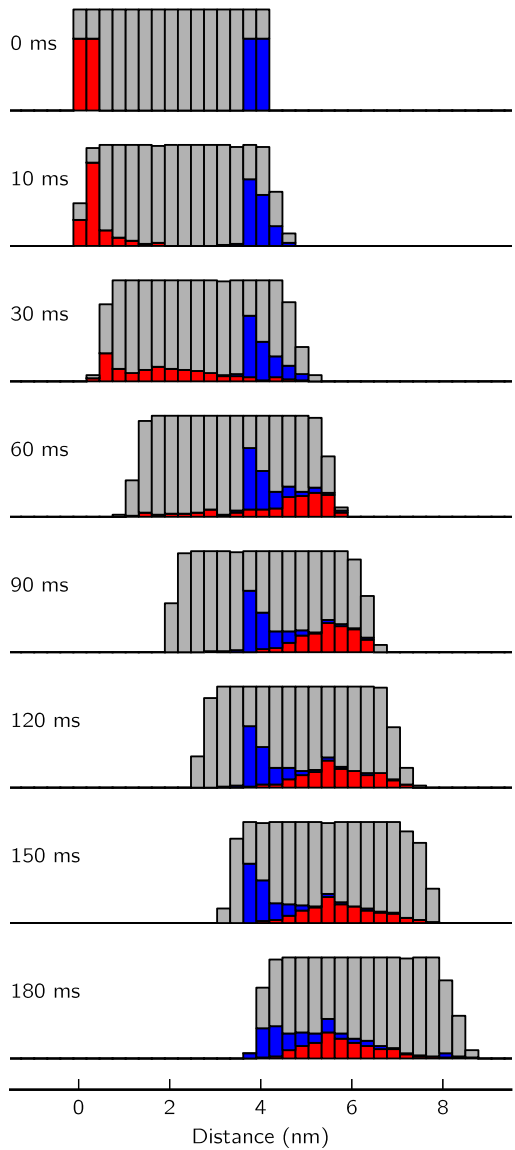}
\caption{Results from the same kinetic Monte Carlo simulation as in
  Fig.~\ref{fig:straight_occ} in the form of a histogram indicating
  the number of iron atoms within each bin with length of one lattice
  constant($a=0.29$~nm). The number of atoms initially in region A (B)
  are colored red (blue) and their positions are tracked during the
  simulation. Vertical position of the bars in the histograms are
  arbitrary, only the height of each individual bar (gray, red, or
  blue) is to be interpreted as the number of atoms within that bin.}
\label{fig:straight_profile}
\end{figure}

Figures~\ref{fig:straight_occ} and \ref{fig:straight_profile} show
eight snapshots of the single kinetic Monte Carlo simulation of the
iron nanocrystal movement inside a carbon nanotube with constant
cross-section. The first snapshot ($t=0$~ms) corresponds to the
initial configuration, the second snapshot follows after $t=10$~ms,
while the remaining six snapshots all follow in intervals of 30~ms
from the initial configuration. Figure~\ref{fig:straight_occ} shows
projection of atom coordinates (gray spheres) onto two-dimensional
plane parallel to the cylinder (nanotube) axis. From this figure we
can see that the carbon nanotube in this particular configuration
moves by its one length in roughly 180~ms.


Figure~\ref{fig:straight_profile} shows, for the same kinetic Monte
Carlo run as in Fig.~\ref{fig:straight_occ}, the distribution of atom
occupation in the form of a histogram. Each bin in the histogram has a
length of one lattice constant ($a=0.29$~nm), and its height
represents the number of iron atoms within that region of the
nanocrystal. Additionally, Fig.~\ref{fig:straight_profile} indicates,
in red and blue, the number of atoms that are in the initial
configuration ($t=0$~ms) in regions A and C respectively (vertical
position of gray, red, and blue regions is meaningless). In subsequent
snapshots, these atoms move from one region to another, as discussed
previously.  For example, we find that atoms which at $t=0$~ms are in
region A (red) by time $t=30$~ms are almost entirely in region B.  By
$t=90$~ms these atoms are distributed along regions C and D, while at
$t=180$~ms they are entirely in region D. On the other hand, atoms
initially ($t=0$~ms) in region C (blue) are in region D by $t=30$~ms
and remain stationary in region D until entering region A at
$t=180$~ms.

\begin{figure}
\centering\includegraphics{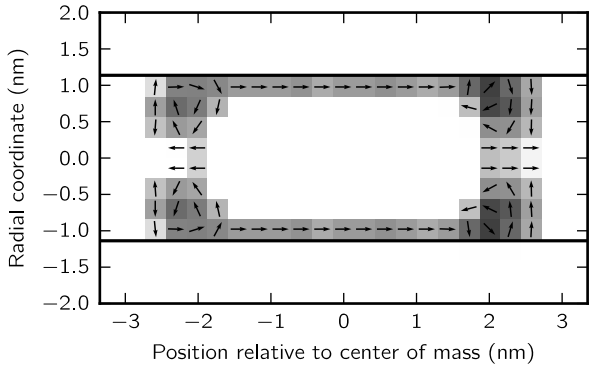}
\caption{Flow of iron atoms in the iron nanocrystal during its
  movement through the nanotube. See main text for a more detailed
  description. Larger flow magnitude is indicated with darker shade of
  gray, on a logarithmic scale. Direction of flow is shown with arrows
  (regardless of magnitude, all arrows have the same length). For
  clarity all flow information is duplicated from positive radial
  coordinates to the negative coordinates. The simulation temperature
  in this calculation is 600~K, somewhat larger than in the
  experiment\cite{Begtrup2009}, iron nanocrystal radius is
  $r_{\textrm{cyl}}$=1.11~nm, and the length is $l=4.31$~nm. The
  electromigration force magnitude is $F_{\rm em}=0.28$~eV/nm.}
\label{fig:flow}
\end{figure}

Finally, Fig.~\ref{fig:flow} shows the computed flow of atoms in the
nanocrystal as a function of their radial coordinate and the
coordinate along the cylinder axis. For each step in the kinetic Monte
Carlo simulation we recorded the initial atomic coordinate (${\bm
  r}_i$) in the nanocrystal center-of-mass reference frame and
direction of atomic step (${\bm r}_j-{\bm r}_i$). For a given
coordinate ${\bm r}_i$ averaging the directions of performed atomic
steps over all kinetic Monte Carlo steps involving site ${\bm r}_i$
gives us a flow vector ${\bm f}_i \sim \sum ({\bm r}_j-{\bm r}_i)$ at
that point. Darker regions in Fig.~\ref{fig:flow} indicate points with
larger magnitude of flow vector ${\bm f}_i$ in logarithmic scale.
Arrows in Fig.~\ref{fig:flow} indicate the direction of flow vector
${\bm f}_i$.  We have neglected azimuthal components of ${\bm f}_i$.
Additionally, Fig.~\ref{fig:flow} shows flow vectors summed over
azimuthal component of initial coordinate ${\bm r}_i$.

We conclude from Fig.~\ref{fig:flow} once again that atoms are moving
only on the surfaces (regions A, B, and C) while they remain
stationary in the bulk (region D). Furthermore, from here we infer
that atomic flow in regions A and C (caps) is about 10 to 100 times
larger than flow in region B (this difference is somewhat obscured by
the logarithmic scale in Fig.~\ref{fig:flow}).

\subsubsection{Dependence on nanocrystal length}
\label{sec:length}

In our kMC simulations we varied nanocrystal lengths from $l=3$~nm up
to $l=40$~nm (in temperature ranges from 500 to 900~K). We find that
the center of mass speed is nearly independent of the nanocrystal
length.  This means that movement of iron atoms near the carbon
nanotube (region B) is much more efficient than the diffusion in
regions A and C. In other words, it takes an iron atom long time to go
from region A to B (or vacancy to go from B to C), but once iron atom
reaches region B it proceeds quickly to region C on the other side of
the nanocrystal.

\subsubsection{Dependence on temperature and electromigration force}
\label{sec:tempForce}

We find a very strong dependence of the nanocrystal center of mass speed on
the temperature and electromigration force. Circular symbols in
Fig.~\ref{fig:extrapolate} show kinetic Monte Carlo results for the iron
nanocrystal center of mass speed on a logarithmic scale for varying
temperature and electromigration force. (Similarly,
Fig.~\ref{fig:exp_regime} shows in linear scale speed for a single
temperature.) Nanocrystal cross-sectional area and length in this
calculation are kept constant.

When the electromigration force on iron atoms becomes too large, we
find that the iron nanocrystal movement becomes unstable and it can
breakup into smaller pieces. Occurrence of such instability in the
model also depends on the thickness of region in which iron atoms
experience electromigration force, and we discuss this dependence in
more detail in Sec.~\ref{sec:robustness}.  Some experimental evidence
for this kind of behavior has been seen in
Ref.~\onlinecite{Svensson2004}.

Kinetic Monte Carlo results shown in Fig.~\ref{fig:extrapolate}
clearly show that motion of iron nanocrystal is temperature activated,
which motivated us to model its movement with that of an effective
single particle in an external potential. In appendix~\ref{app:dif} we
derived an expression for the speed of one particle in periodic
external potential (barrier height $B$ and period $L$) under the
influence of constant external force $F$, and in contact with a
thermal bath at temperature $T$. Using this expression we can now try
to fit our kinetic Monte Carlo results for center of mass speed to the
following functional form,
\begin{align}
  v = \tilde{v} \exp \left( - \frac{ \tilde{B}}{k T} \right) \sinh
  \left( \frac{\frac{1}{2} \tilde{L} F}{k T} \right).
\label{eq:fit}
\end{align}
Here $\tilde{v}$, $\tilde{B}$, and $\tilde{L}$ are fitting parameters
which correspond respectively to the velocity prefactor, barrier
height and period of external potential for this single effective
particle. We set force $F$ to equal electromigration force experienced
by a single iron atom in the simulation, $F_{\rm em}$.

Lines in Fig.~\ref{fig:extrapolate} show the fit of the kinetic Monte
Carlo simulation results to the functional form given by
Eq.~\ref{eq:fit}. One can see that this functional form reproduces
quite well simulated results. Fitted values of for velocity prefactor
$\tilde{v}$, effective barrier height $\tilde{B}$, and effective
period $\tilde{L}$ are,
\begin{align}
  \tilde{v} = 3.3 \textrm{~m/s,}
  \quad 
  \tilde{B} = 1.2 \textrm{~eV,}
  \quad
  \tilde{L} = 1.4 \textrm{~nm.}
\label{eq:vBL}
\end{align}

\subsubsection{Dependence on nanocrystal cross-section area}
\label{sec:area}

Finally, we analyze the dependence of iron nanocrystal center of mass
speed on the nanocrystal cross-section area. The number of atoms that
need to travel from region A to region C in order for the crystal to
move a certain fixed length is proportional to nanocrystal
cross-section area $\sim r_{\textrm{cyl}}^2$. However, with increasing
cross-sectional area the number of pathways to travel through region B
is also increasing, but only as $\sim r_{\textrm{cyl}}^1$. Naively,
one would therefore expect that center of mass speed of an iron
nanocrystal will be proportional to $\sim r_{\textrm{cyl}}^{-1}$.
However, our calculations find that there is lot of variations on top
of overall trend of decreasing center of mass speed with radius
$r_{\textrm{cyl}}$.
The reason for this discrepancy we find in the following. Nature of
diffusion pathways in region B of iron nanocrystal will depend
strongly on the details of the cylindrical boundary of the iron
nanocrystal. For example, we find that for some specific values of
nanocrystal radius $r_{\textrm{cyl}}$ one can have in region B
precisely two rows of iron atoms on top of iron (011) surfaces. Our
model from Eq.~\ref{eq:asym_barrier} predicts that there is very small
diffusion barrier for movement along these two rows of atoms (since
$Z^{i}_{\rm Fe}=Z^{j}_{\rm Fe}=3$) which means that movement along
region B (and possibly into or out of region B) is greatly enhanced.

Repeating the fit to the effective particle model from
Eq.~\ref{eq:fit} for nanocrystals with varying cross-sectional area we
find that fitting parameters appearing in exponential and sinus
hyperbolic functions: $\tilde{B}$ and $\tilde{L}$ are almost
unaffected. Only parameter which seems to depend on cross-section area
is velocity prefactor $\tilde{v}$, which is of smaller importance. For
example, when comparing our results to experiment in
Sec.~\ref{sec:exp_constant} precise value of $\tilde{v}$ will be of
almost negligible importance as compared to values of $\tilde{B}$ and
$\tilde{L}$ appearing inside exponential and sinus hyperbolic
functions in fitting function, Eq.~\ref{eq:fit}.

More specifically, we performed calculations for five different
nanocrystal radii $r_{\textrm{cyl}}$ ranging from $1.05$~nm to
$1.73$~nm, corresponding to cross-sectional area from 3.46~nm$^2$ to
9.40~nm$^2$. Among these five calculations we find that largest fitted
value of parameter $\tilde{v}$ is about three times larger than for
the case with smallest value of $\tilde{v}$. On the other hand,
parameters $\tilde{B}$ and $\tilde{L}$ are varying only about 10\%.

\subsubsection{Comparison with experiment}
\label{sec:exp_constant}

\begin{figure}
\centering\includegraphics{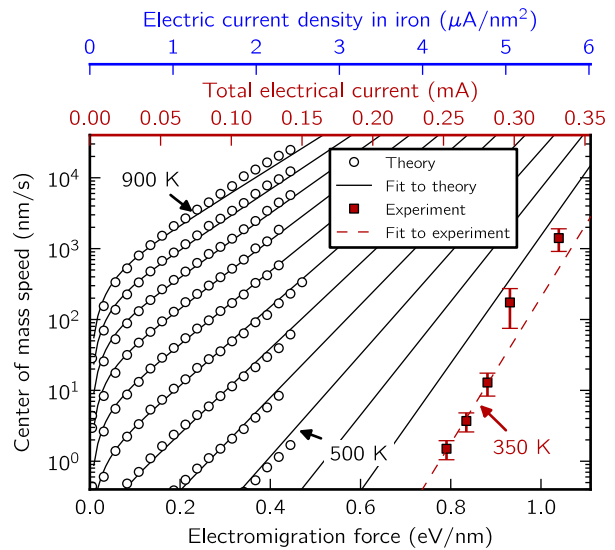}
\caption{Dependence of iron nanocrystal center of mass speed on
  electromigration force (black) and total electrical current $I$
  (red). Kinetic Monte Carlo simulation (black circles) was done for
  temperatures from 500~K to 900~K in steps of 50~K, iron nanocrystal
  radius for all calculations equals $r_{\textrm{cyl}}$=1.05~nm, while
  length is $l=4.31$~nm. A fit to model from Eq.~\ref{eq:fit} with
  parameters given in Eq.~\ref{eq:vBL} is shown with black lines for
  temperatures from 350~K to 900~K in steps of 50~K.  Experimental
  data from Ref.~\onlinecite{Begtrup2009} is shown with red squares,
  and fit to model from Eq.~\ref{eq:fit} to experimental data is shown
  with red dashed line (temperature used in fit is $350$~K, consistent
  with an independent experimental estimate). Relationship between
  electromigration force (bottom axis, black) and estimated current
  density $j$ through the iron nanocrystal (topmost axis, blue) is
  given by Eq.~\ref{eq:Kj} as discussed in the text.}
\label{fig:extrapolate}
\end{figure}

In Ref.~\onlinecite{Begtrup2009} the speed of an iron nanocrystal was
measured as a function of an applied external voltage $V$ and current
$I$ (red square symbols in Fig.~\ref{fig:extrapolate}). On the other
hand, in our kinetic Monte Carlo simulation we compute the speed of an
iron nanocrystal as a function of electromigration force $F_{\rm em}$
(black circles in Fig.~\ref{fig:extrapolate}).  In order to relate
$F_{\rm em}$ to $I$ we first assume that the electromigration force
$F_{\rm em}$ is linearly proportional to the current density $j$,
\begin{align}
  F_{\rm em} = K j,
  \label{eq:Kj}
\end{align}
and we later obtain the parameter $K$ by fitting to the experiment.
(The linear dependence of $F_{\rm em}$ on $j$ as in Eq.~\ref{eq:Kj} is
consistent with an electron wind force mechanism as discussed in
Refs.~\onlinecite{Sorbello1997,Dekker1998}.)

We crudely estimate the current density $j$ in the iron nanocrystal by
making the following set of assumptions. First, we assume a constant
current density profile perpendicular to the carbon nanotube axis,
both in the iron nanocrystal and in the carbon nanotube. Second, we
assume that the conductivity of the iron nanocrystal equals that of
the bulk iron. Both of these assumptions likely underestimate the
current density $j$ (and therefore overestimate $K$). Nevertheless,
under these assumptions, current density $j$ flowing through the iron
nanocrystal is given as,
\begin{align}
  j = \frac{I}{A_{\rm tube}} \left( \frac{\rho_{\rm iron}}{\rho_{\rm
        tube}} + \frac{A_{\rm iron}}{A_{\rm tube}} \right)^{-1}.
\label{eq:estimJ}
\end{align}
Here, $A_{\rm tube}$ and $A_{\rm iron}$ are cross-sectional area of
carbon nanotube and iron nanocrystal respectively. We estimate $A_{\rm
  tube}$ and $A_{\rm iron}$ from inner and outer diameters of the
carbon nanotube used in Ref.~\onlinecite{Begtrup2009} ($5-7$~nm and
$35$~nm respectively). For the resistivity of iron $\rho_{\rm iron}$,
we use $8.6 \cdot 10^{-8}$~$\Omega$~m, while the resistivity of the
carbon nanotube $\rho_{\rm tube}$ we can compute from the length of
the tube (2~$\mu$m), $A_{\rm tube}$, $V$, and $I$. This procedure
gives us $\rho_{\rm tube} \sim 2.6\cdot10^{-6}$~$\Omega$~m, close to
the bulk resistivity of graphite.

We obtain good agreement with experimental
measurements\cite{Begtrup2009} of the iron nanocrystal center of mass
speed using $K=0.18$~eV~nm/$\mu$A and temperature $T=350$~K (compare
dashed red line and red symbols in Fig.~\ref{fig:extrapolate}).
However, we expect that there is a large uncertainty in value of
parameter $K$ due to our crude estimate of current density $j$. We are
unaware of any other theoretical or experimental estimates of
electromigration force coefficient $K$ in iron. Additionally, the
value of the parameter $K$ varies a lot across the periodic
table\cite{Dekker1998} both in magnitude and sign.  Furthermore, the
value of the parameter $K$ is very sensitive to the structural
parameters. For example, it can vary a great deal between fcc and bcc
phases of the material\cite{Dekker1998}. Interestingly enough, the
largest value of the parameter $K$ among all cases studied in
Ref.~\onlinecite{Dekker1998} is that of an iron impurity
electromigrating in aluminum ($K=0.01$~eV~nm/$\mu$A), which is within
an order of magnitude of our estimated value of $K$.

Experiments in Ref.~\onlinecite{Begtrup2009} have been performed at
room temperature, but the actual temperature on the carbon nanotube
has not been measured directly. Independent estimates, based on Joule
heating and thermal conductivity of the Si$_3$N$_4$ substrate, give an
estimated temperature of $\approx 440$~K, consistent with our fitted
value.  A similar value is obtained by scaling the Joule heating power
to that used in Ref.~\onlinecite{Begtrup2007} where temperature of the
carbon nanotube has been directly measured.

\subsubsection{Robustness of results on model parameters}
\label{sec:robustness}

Now we will discuss robustness of our results on changes in model
parameters. There are four parameters ($a$, $b$, $c$, and $d$) in
Eq.~\ref{eq:asym_barrier} which have all been fitted to
first-principles DFT calculation. Additionally we assumed that the
electromigration force ${\bm F}_{\rm em}$ influences only iron steps
when either initial site $i$, or final site $j$ are immediately next
to the carbon nanotube.

Let us start by testing robustness of our results on four parameters
from Eq.~\ref{eq:asym_barrier}. We performed series of calculations in
which we either increased or decreased by 15\% each of these four
parameters separately. We find in all eight calculations that
qualitative character of iron nanocrystal movement remains unchanged.
Additionally, dependence on temperature and electromigration force
remains qualitatively the same as in Eq.~\ref{eq:fit}. Quantitatively,
we find small changes in the fitting parameters $\tilde{v}$,
$\tilde{B}$, and $\tilde{L}$. The resulting iron nanocrystal center of
mass speed is more sensitive to parameters $b$ and $c$ than to $a$ and
$d$.

Additionally, we tried removing the dependence of diffusion energy
barrier height on initial number of first neighbor iron atoms $Z_{\rm
  Fe}^i$.  Therefore we set parameter $b$ to zero and vary value of
parameter $a$. We again find qualitatively the same dependence of
center of mass speed as in Eq.~\ref{eq:fit}. We changed the value of
the parameter $a$ from 0.4 to 0.7~eV and the main quantitative
difference we find is that effective period $\tilde{L}$ is about two
times smaller then using original values of $a$, $b$, $c$, and $d$
parameters.

Another robustness test we performed is to increase region in which
iron atoms feel influence of the electromigration force ${\bm F}_{\rm
  em}$. Instead of just considering atoms which are in contact with
carbon atoms, we redid calculation in which this region was increased
so as to include iron atoms up to 0.4~nm away from the carbon
nanotube. Also, as an extreme case, we redid calculation where
electromigration force ${\bm F}_{\rm em}$ was acting on all iron
atoms. We find that with different regions in which the force ${\bm
  F}_{\rm em}$ is acting on the iron nanocrystal center-of-mass is
almost unaffected.

Nevertheless, we find that with increasing region in which force ${\bm
  F}_{\rm em}$ is acting, iron nanocrystal starts to breakup at
smaller and smaller forces. When only first layer of iron atoms is
experiencing electromigration force, nanocrystal starts to break when
force $|{\bm F}_{\rm em}|$ is larger than 0.5~eV/nm (other parameters
are as in data from Fig.~\ref{fig:exp_regime}). When iron atoms up to
0.4~nm away from the nanotube are experiencing electromigration force,
nanocrystal breaks up for forces above 0.25~eV/nm. Finally, when all
iron atoms are experiencing electromigration force, breaking occurs
already above 0.15~eV/nm.

\subsection{Movement through a constriction}

\begin{figure}
\centering\includegraphics{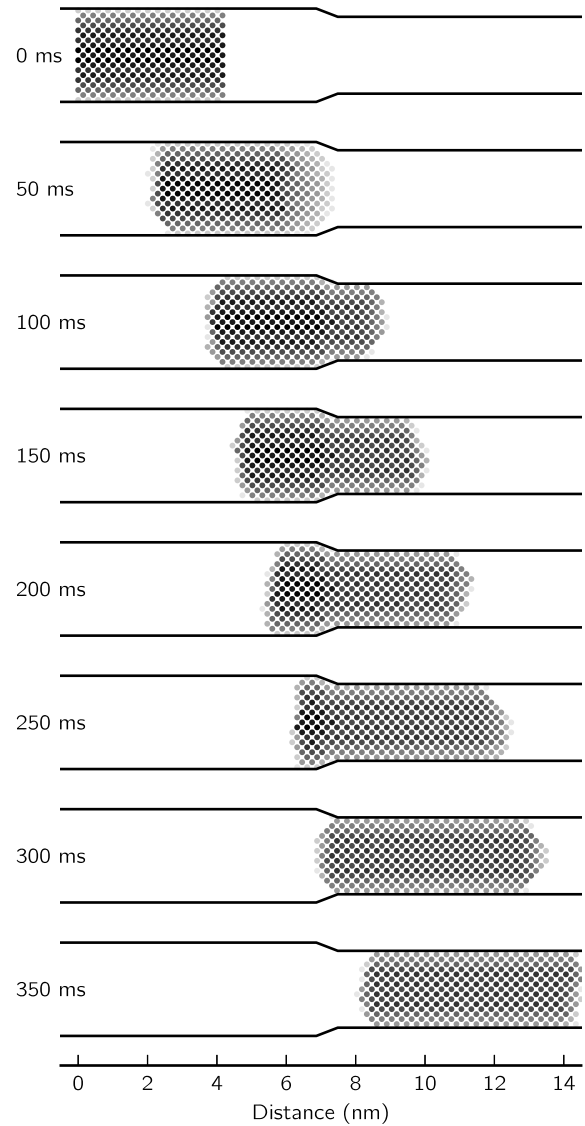}
\caption{Two dimensional projection of three dimensional iron atom
  positions at eight different snapshots in the kinetic Monte Carlo
  simulation, as in Fig.~\ref{fig:straight_occ}. Various intensity of
  greyness correspond to rows containing more (darker gray) or less
  (brighter gray) iron atoms. The simulation temperature is 700~K,
  somewhat larger than in the experiment\cite{CohPRL2013}, iron
  nanocrystal radius in the region on the left is
  $r_{\textrm{cyl}}$=1.35~nm, and on the right is
  $r_{\textrm{cyl}}$=1.11~nm (as in Fig.~\ref{fig:straight_occ}).
  Therefore, cross-section area on the left is about 50\% larger than
  on the right. Length of iron crystal at $t=0$~ms is $l=4.31$~nm. The
  electromigration force magnitude is $F_{\rm em}=0.28$~eV/nm.}
\label{fig:constriction}
\end{figure}

Now we will describe movement of iron nanocrystal through a tube with
a varying cross-section. At first, it seems surprising that solid
piece of iron nanocrystal could move through constrictions in nanotube
with cross-section smaller than nanocrystal cross-section. However,
taking into account the character of the iron nanocrystal movement we
discuss in Sec.~\ref{sec:character}, it becomes clearer why this is
possible.  Iron atoms in region D remain stationary and therefore do
not need to move through a constriction directly. On the other hand,
when iron atoms move from region B into region C, they adapt to tube
with smaller cross-section. Movement of iron nanocrystal through a
constriction in the carbon nanotube has been experimentally
demonstrated in Ref.~\onlinecite{CohPRL2013}.

Figure~\ref{fig:constriction} shows kinetic Monte Carlo simulation
results of a movement of iron nanocrystal through a tube with area
5.7~nm$^2$ that gets shrunk to 3.9~nm$^2$. One can see that between
moment $t=50$~ms and $t=300$~ms iron nanocrystal was able to move
through a constriction. We find the same behavior for other ratio of
nanotube cross-sections.

\section{Conclusion}

Our first-principles based on kinetic Monte Carlo simulations of iron
nanocrystal inside carbon nanotube show the nature of movement of iron
nanocrystal. We find that the iron nanocrystal does not move as a
whole but instead atoms are moving only on the surfaces, from one end
of crystal to the other. See Sec.~\ref{sec:character} for more
detail. Consistent with this observation, we also find that an iron
nanocrystal is able to move through a constriction in the carbon
nanotube that has larger diameter than the nanocrystal.

Somewhat surprisingly we find theoretically that an iron nanocrystal
center of mass speed does not depend on the length of the nanocrystal.
We attribute this to the fact that individual iron atom moves through
region B quite fast, compared to time spent in region A, or C.
Furthermore, we find that movement of entire nanocrystal can be
modeled quite well as thermally activated motion of single particle in
tilted periodic potential with period of 1.4~nm, and barrier height
1.2~eV, regardless of carbon nanotube area, length, temperature, or
electromigration force. In future, it would be interesting to measure
experimentally dependencies of center of mass speed on nanocrystal
length, area, and temperature. So far, only dependence on external
current has been established, for fixed length, area, and temperature.

In our model we assumed that only iron atoms next to the carbon
nanotube are experiencing electromigration forces. Nevertheless, even
if we allow a larger region of iron atoms to experience electromigration
force (or even entire iron nanocrystal) we still find that iron
nanocrystal can move through the carbon nanotube. However, as this
region gets larger and larger, movement of iron nanocrystal becomes
more and more unstable.

Comparing the experimentally measured speed of an iron nanocrystal with our
model calculation we estimate that temperature of iron nanocrystal is
not much larger than room temperature ($\sim350$~K) which is in
agreement with crude estimates from Joule heating. Furthermore, we find
that relationship between current density through iron nanocrystal and
force on individual iron atoms is given by constant of proportionality
$K=0.18~\textrm{eV nm}/\mu\textrm{A}$.

\begin{acknowledgments}
  We thank David Strubbe for discussion. This work was supported by
  the Director, Office of Energy Research, Office of Basic Energy
  Sciences, Materials Sciences and Engineering Division, of the U.S.
  Department of Energy under Contract No. DE-AC02-05CH11231.
\end{acknowledgments}

\appendix
\section{Diffusion in one-dimensional periodic potential}
\label{app:dif}

Diffusion in a one-dimensional periodic potential $U(x+L)=U(x)$ under
application of an external force $F$ can be modeled by the following
equation of motion,
\begin{align}
  \eta \frac{d x}{ dt} = F - \frac{d U}{d x} + \sqrt{2 \eta k T}
  \xi(t).
\end{align}
Here $\eta$ is friction coefficient, and $x(t)$ is position of
particle at time $t$. The stochastic force on the particle due to
thermal fluctuations at temperature $T$ is modeled by a random
variable $\xi(t)$ with zero mean value, $\left\langle \xi(t )\right
\rangle=0$ and a Dirac delta correlation, $\left\langle \xi(t )
  \xi(t') \right \rangle = \delta(t-t')$. The analytic expression for
the average velocity of the particle governed by such equation is
given as\cite{Risken1996},
\begin{align}
  v = \frac{L k T}{\eta} \frac{ 1 - \exp\left(-\frac{L F}{k T}\right)}
  { \int_0^L \int_0^L \exp \frac{ U(x) - U(x-x') - F x' }{kT} dx dx'
  }.
\end{align}
For a sawtooth potential ($U(x)=x \frac{2 B}{L}$ for $0<x<L/2$ and
$U(x)=2 B - x \frac{2 B}{L}$ for $L/2<x<L$) with period $L$ and
barrier height $B$ one can show that in the limit of $k T \ll B$ and
$\frac{1}{2} F L < B$ velocity of particle is given as,
\begin{align}
  v \approx \frac{2 B^2}{\eta L k T} \exp \left( -\frac{B}{k T} \right)
  \sinh \left(\frac{\frac{1}{2} L F}{k T} \right).
\end{align}

\bibliography{pap}

\end{document}